# A physics inspired and efficient transform for optoacoustic systems

Authors: María Rodríguez Sáenz de Tejada[1,2], Alvaro Jimenez[1,2], Rodrigo Rojo[1,2], Sergio Contador[1,2], Juan Aguirre[1,2]*

Affiliations:

1. Department of Electronic and Communications Technology, Medical Engineering Development and Innovation Center, Autonomous University of Madrid, Madrid, Spain.

2. Health Research Institute of the Jiménez Díaz Foundation, Madrid, Spain.

*Corresponding author

## Abstract

Optoacoustic imaging technologies require fast and accurate signal pre-processing algorithms to enable widespread deployment in clinical and home-care settings. However, they still rely largely on the Discrete Fourier Transform (DFT) as the default tool for essential signal-conditioning operations, which imposes hard limits on both execution speed and signal-retrieval accuracy.

Here, we present a new transform whose building blocks are directly inspired by the physics of optoacoustic signal generation. We compare its performance with the DFT and other classical transforms on common signal-processing tasks using both simulations and experimental datasets. Our results indicate that the proposed transform not only sets a new lower bound on computational complexity relative to the DFT, but also substantially outperforms classical transforms on basic signal-processing operations in terms of accuracy. We expect this transform to catalyze broader adoption of optoacoustic methods.

## Introduction

In the late 19[th] century Joseph Fourier introduced the Fourier transform as an auxiliary tool for his analytical studies of heat diffusion. Since then, the transform and its digital version, the Discrete Fourier Transform (DFT), has impacted countless fields in science and engineering [1] .

In the field of biomedical imaging the DFT plays a preponderant role for essential signal conditioning procedures. Optoacoustic (OA) imaging systems are not an exception, and rely heavily on the DFT to perform operations like random and non-random electrical noise removal or DC component suppression [2], [3].

Over the past two decades, a wide range of OA imaging implementations have been developed, establishing optoacoustics as a powerful tool for biological research and a promising modality for clinical imaging and home health care. This success stems from OA imaging's unique combination of optical-absorption contrast, high spatial resolution, and deep penetration. Together, these features provide a competitive edge over alternative imaging modalities in oncology, cardiology, dermatology, endocrinology, and other medical fields [4], [5], [6], [7], [8], [9], [10], [11], [12],[13], [14],[15], [16], [17].

OA signals are generated by illuminating tissue with pulsed light. As the light propagates through the tissue, absorbed energy induces thermoelastic expansion, which in turn generates ultrasound waves. Raw OA signals are acquired by recording these acoustic waveforms with ultrasound transducers placed at different positions on the tissue surface. After signal conditioning—typically performed using the DFT—images can be formed.

However, the characteristics of the data sets produced by OA systems often pose a significant computational challenge for the DFT. Generally, OA data sets are multidimensional, combining ultrawideband acoustic signals from a disparity of locations and multiple wavelength data [18]. The most efficient algorithmic form of the DFT (the Fast Fourier Transform, FFT) was introduced in 1965 [19], setting since then a hard limit that remains unbeaten. Even when using the FFT, standard PCs may need several minutes to hours for signal conditioning, which is incompatible with clinical workflows, home-care applications, and even many biological research settings.

Importantly the DFT was derived as a general purpose algorithm intended to be applicable across any kind of technology. In OA, the data processing schemes related to the DFT inevitable loose components from the original signal [2], [3].

Alternative analytical transforms such as the wavelet transform (WT) [20] have been explored in OA imaging [2], [3]. The discrete wavelet transform can be tailored for OA technologies by selecting building blocks that loosely attain the characteristics of optoacoustic signals [21], albeit without achieving a explicit connection with the underlying physics of OA. Moreover, it overcomes the computational complexity of the FFT. On the other hand, modern AI based denoising models are effective at removing specific noise patterns through appropriate training. However, their effectiveness is compromised since they cannot be easily adapted to novel noise sources [22].

In here we hypothesized that by designing a transform (namely the discrete optoacoustic transform DOAT), inspired on the physics of OA signal generation, we can overcome the limits of classical transforms in common OA signal conditioning tasks. Our result show that DOAT drastically overcomes the computational complexity of FFT while also strongly improving accuracy compared with both the FFT and the DWT.

## Results

**Foundations of the DOAT and its computational complexity**

The strength of the DFT for signal processing is rooted on the fact that it is used to decompose any given discrete signal into a sum of sinusoidal functions whose terms can be manipulated at will. Each sine or cosine has a specific frequency and its contribution to the summation is weighted by a corresponding coefficient. Lower frequency sinusoidals capture the coarse structure of the signal while the higher frequencies capture the finer (see **Figure 1a**). By modifying the coefficients, the contribution to the signal of each sinusoidal component can be weighted as desired. Like this, specific features of the signal can be enhanced or attenuated while preserving the remainder. These coefficients, known as Fourier coefficients, are obtained by applying the DFT to the original signal.

Similarly, the DWT is generally used to decompose a signal into of a sum of functions which can be modulated at will by changing the value of their respective coefficients (see **Figure 1b**). As in the Fourier case, each function is related to the coarser or finer details of the signal, however, they can also be shifted in time.

In both the DWT and the DFT, the functional components are referred to as "*basis functions*", and must fulfil several mathematical properties (see Discussion). In the case of the DWT, there exists a wide variety of "*basis families*" each with a different functional form.

The DOAT also decomposes the signal into a sum of functions, however, its basis functions and their combination resemble the physics of OA signal formation. Specifically, by performing the DOAT, it is assumed implicitly that a given OA time signal is the result of the combination of individual time signals corresponding to individual spherical emitters of different sizes (see **Figure 1c**).

An OA signal emitted by a sphere exhibits a characteristic N-shape profile. The sphere diameter is encoded in the temporal separation between the two vertical edges of the N (i.e., $d = \Delta t\, v_s$, where $v_s$ is the speed of sound; see Fig. 1d), whereas the absorbed optical energy is reflected in the peak-to-peak amplitude of the N (see Fig. 1d). Notably, constructing the N-shaped waveform requires only the locations and amplitudes of these two edges. Accordingly, the DOAT basis functions correspond to the vertical edges ("posts") of the N-shape waveform, while the coefficients represent the absorbed optical energy of the sphere (see Methods).

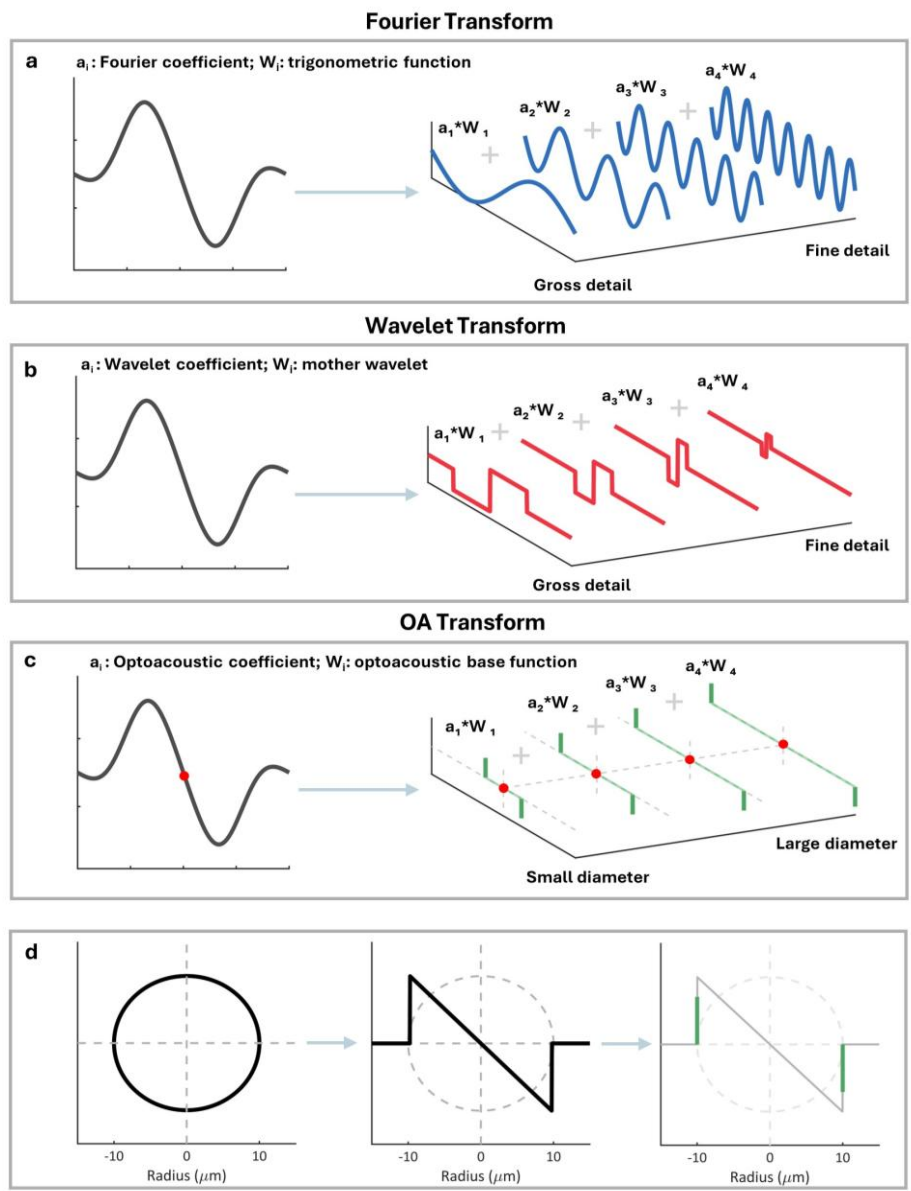

**Figure 1. Depiction of signal decomposition according to different transforms. (a)** DFT decomposition of a signal (left) into a sum of sinusoidal functions (right). **(b)** DWT decomposition of a signal (left) into a sum of basis functions (right). **(c)** DOAT decomposition of a signal (left) into a sum of basis functions (right). Marked in red, the known center of the signal (left), which then corresponds to the center of the basis functions (right), also marked in red. **(d)** Schematics of a 10μm radius sphere (left), the signal generated by the sphere (center) and the DOAT basis function related to the said signal, marked in green (right).

While the DFT, in its most efficient implementation (FFT) has a computational complexity of $Nlog(N)$ the DOAT has a computational complexity of $N$. This reduction arises because the DOAT basis function are sparse (all zeros except two points). As a result, computing the transform requires far fewer operations to obtain the coefficients (see Methods).

As in the DFT and the DWT the DOAT basis functions fulfil several mathematical properties that justify the applicability of the transform (see Discussion).

**Accuracy of the DOAT vs classic transforms in simulated data.**

We wanted to study the performance of the transform for two of the main signal processing operations: random and non-random electrical noise removal and DC component removal. To this end, we simulated ideal optoacoustic time signals from spheres of biologically relevant sizes, added the effects of a transducer Electrical Impulse Response (EIR, non-random electrical noise), a large amount of random noise (random electrical noise) and a DC component (see Methods). We then applied the DFT, DWT and OAT to perform denoising and compared the outputs with the ideal signals.

**Figure 2a** shows an ideal optoacoustic signal corresponding to a sphere of 10µm radius. **Figure 2b** represents the signal after applying the impulse response of a transducer (see Methods) commonly used in optoacoustic imaging for spheres of the said sizes. Then the DC component and random electrical noise is added (see Methods and **Figure 2c**).

After computing the DFT, DWT and the DOAT on the signals we obtained the coefficients of each transform (**Figure 2d,e** and **f**). We then filtered these coefficients (**Figure 2g,h** and **i**) and calculated the inverse transform for each method, obtaining the processed signal (**Figure 2j,k,l**).

Regarding the DFT we used a 4[th] order Butterworth filter between 10MHz and 120MHZ, which is a common procedure in optoacoustic signal processing. It is a band-pass filter that on one hand removes the high frequency random noise and the low frequency DC noise component. The upper and lower frequency limit of the filter corresponds to the detection frequency of the simulated transducer (see Methods).

For the DWT we used different sets of mother wavelets and filtering schemes following previous work found in the literature (see methods).

Regarding the DOAT we filtered the coefficient as follows: first we set to zero all coefficients with negative values. By removing negative values, we assume that no change in the polarity of the optoacoustic waves occur. We also set to 0 the coefficients of basis function of spheres of radius larger than 100µm. 100µm is the lower resolution limit according to the transducer bandwidth. Then we applied a moving average of length 150 to all the coefficients to account for the random noise (see Methods).

A quantitative comparison with the original signals (see **Figure 2m** and Methods) clearly indicates that the DOAT overcomes the accuracy of both DFT and DWT, removing effectively the random and non-random noise together with the DC component.

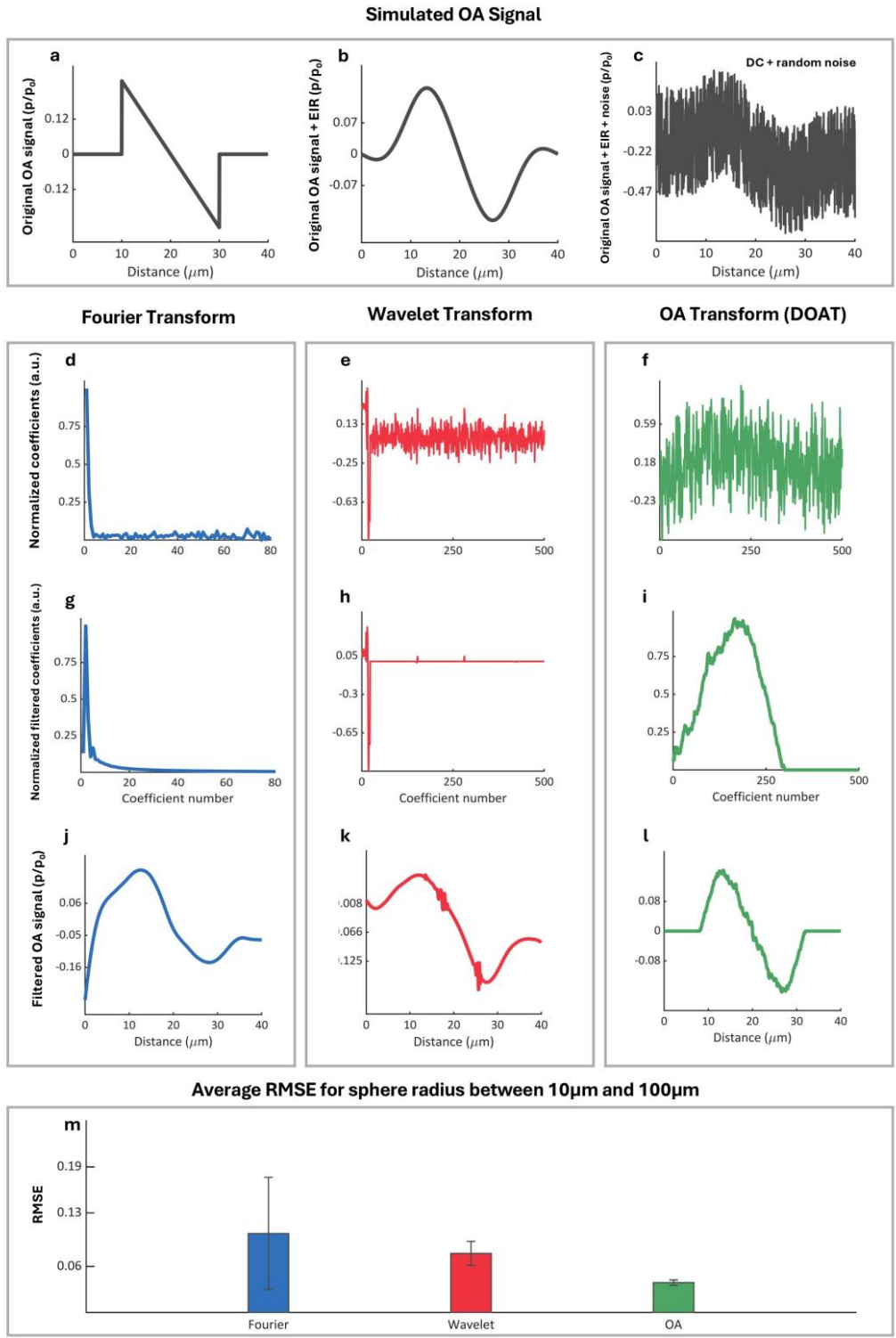

**Figure 2. Study of the performance of the DFT, DWT and DOAT over simulated optoacoustic signals.** **(a)** Ideal optoacoustic signal corresponding to a sphere of 10μm radius. **(b)** Optoacoustic signal convoluted with the impulse response of a 10MHz-120MHz transducer. **(c)** Optoacoustic signal with transducer impulse response, DC noise and high frequency noise. The figure shows the coefficients of the signal after being decomposed by **(d)** DFT (single sided)**, (e)** DWT and **(f)** DOAT. Also shown are the coefficients filtered by **(g)** DFT, **(h)** DWT and **(i)** DOAT. The processed signals are shown, after being filtered with **(j)** DFT, **(k)** DWT and **(l)** DOAT. **(m)** Average RMSE value for each transform with signals corresponding to spheres with radii between 10μm and 100 μm

**Accuracy of the DOAT vs classic transforms in real imaging data.**

**DOAT in optoacoustic microscopy.**

We wanted to test the ability of the DOAT to process real data, in particular optical resolution optoacoustic microscopy data (OR-OAM). OR-OAM systems are gaining traction for major applications in diabetes, cardiovascular disease, spectroscopy and are revolutionizing biological research [4],[23],[24].

In OR-OAM systems a beam of light is focused onto the sample which generates an optoacoustic signal corresponding to the focal zone of the beam. If the focused region contains enough absorbing moieties the signal will display an individual *N-shape*, corrupted by noise. Otherwise the signal will contain only noise. By performing a point-by-point illumination of the sample, a set of signals is obtained (A-lines) from which an image can be formed.

To validate the DOAT data processing capabilities, we acquired real optoacoustic signals by imaging a USAF resolution target using a custom OR-PAM (**see supplementary Note 1 and Figure 3a and 3b**).

We processed the acquired data using both the Discrete Fourier Transform (DFT) and the Discrete Optoacoustic Transform (DOAT) for noise removal. The raw and filtered signals corresponding to a representative A-line are displayed in **Figure 3c**. For the DFT, we applied a 4th-order Butterworth bandpass filter (10–120 MHz) to match the transducer's operational bandwidth. For the DOAT, we eliminated negative values and zeroed all coefficients corresponding to spherical basis functions with radii larger than 100 μm, applying a 10-point moving average to the remaining coefficients. Simple visual inspection of the signals clearly demonstrate the superiority of the DOAT transform.

The resulting processed images are shown in **Figure 3d**. An analysis of the Signal-to-Noise Ratio (SNR) indicates that the DOAT outperforms the DFT (**Figure 3e and methods**). Additionally, we calculated the Root Mean Square Error (RMSE) for both transforms relative to the ground-truth photograph (see methods) shown in **Figure 3b**. Again, simple visual inspection of the images clearly demonstrate the superiority of the DOAT transform.

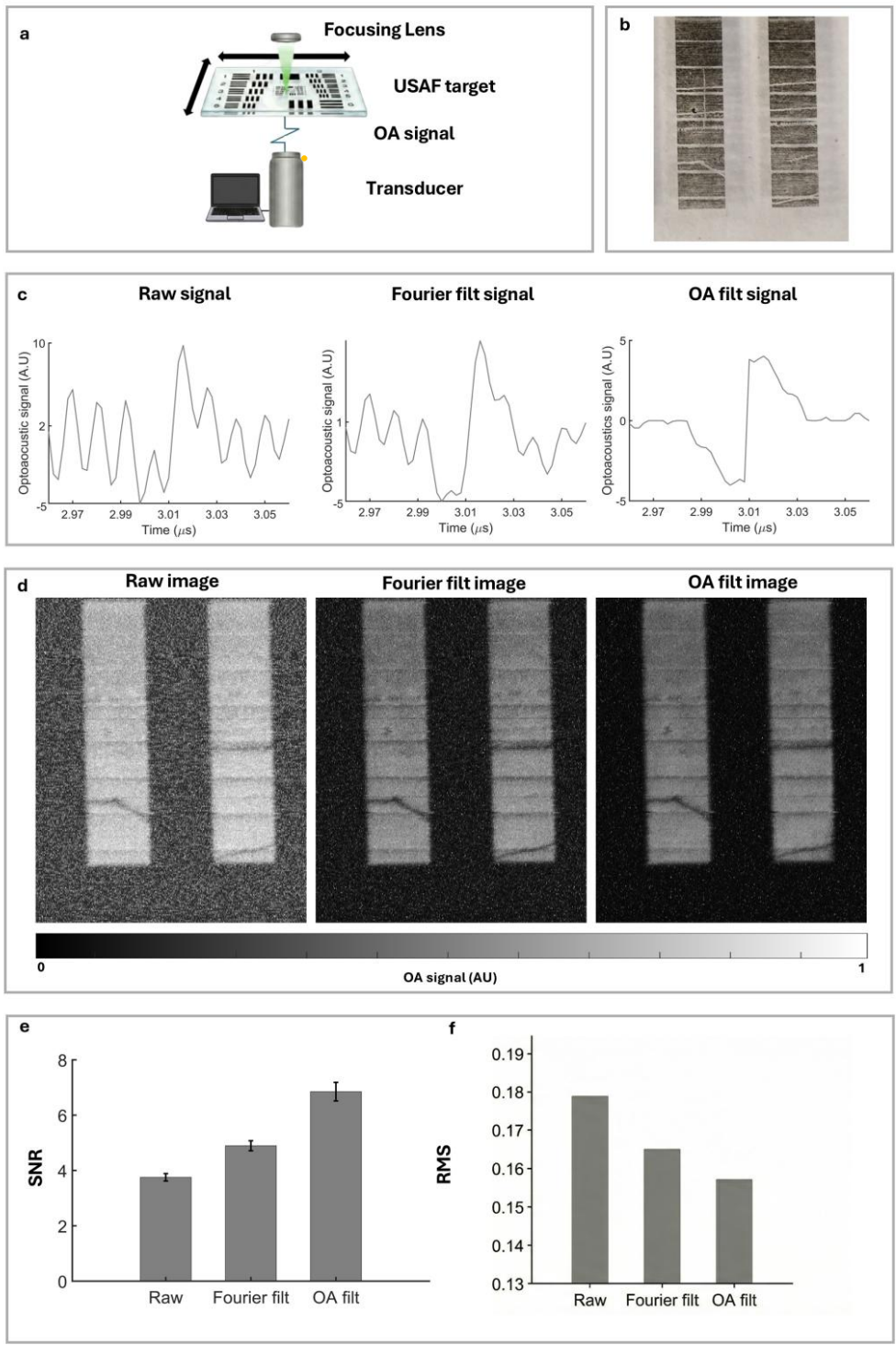

**Figure 3. Study of the performance of the DFT and DOAT over a optoacoustic imaging data. (a)** Schematics of an optoacoustic microscopy system, where the transducer is fixed and collects the emitted OA signal of the scanned target by the laser. **(b)** Real image of the USAF. **(c)** The raw, DFT filtered and DOAT filtered signals. **(d)** The resulting filtered images from the maximum intensity projection of the Raw, DFT and DOAT signals. **(e)** The SNR value of the different resulting images. **(f)** The RMS error of the images when compared to the ground-truth photograph at three different sites of the images.

**DOAT in optoacoustic tomography.**

The DOAT algorithm needs a modification to be used in optoacoustic tomography systems. This is due to the fact that in such systems the OA signals generated at different depths can overlap, therefore, there may exist multiple mixed *N-shapes* rather than one in the resulting signal. In this case the DOAT has to be applied using a moving window similarly as it is done in short-time Fourier transform [25] . We termed this algorithm the SDOAT (see Discussion and Methods).

## Discussion

In here we show a new transform tailored specifically for OA signal processing which we termed DOAT. DOAT has important advantages over classical transforms that arise from the fact that its basic components are inspired in the physics of OA signal formation. First, it breaks through the computational complexity of DFT. Second, it displays a better performance for essential signal processing operations related to noise removal.

From a pure mathematical perspective, the transform presented here is tailored for discrete finite signals and ideally should conform a Hamel basis. OA basis functions are orthonormal which is one of the properties that a Hamel basis should fulfil. However, for signals of length L, the number of basis functions is (L-1)/2 and we need to have L elements. The rest of the elements needed to complete a Hamel basis can be easily found by means of the Kramer method (data not shown). Further work should investigate if the transform can also be generalized for the continuous and infinite case, conforming a Hilbert basis.

The DOAT represents a clear breakthrough in computational complexity in comparison with the DFT for optical resolution optoacoustic microscopy. However, one may argue that the computational complexity of the overall signal process algorithm is not advantageous. This is not the case for the filtering scheme presented in here, since the moving average algorithm has a computational complexity of $N + M$ where $N$ is the size of data and $M$ the number of elements of the moving average

The SDOAT showed a very good performance for bandpass filtering operations in optoacoustic tomography data (data not shown). However, in its current implementation, it loses its computational complexity advantage over DFT and DWT (the computational complexity of SDOAT is $MNlog(N)$). Further work should consist in optimizing the computational complexity of SDOAT. Further work should also investigate what the SDOAT can bring to optoacoustic tomography systems. We foresee possible applications in segmentation, automatic size measurement or even super-resolution methods.

With the advent of optoacoustic imaging systems for relevant applications in the clinics and home care, fast signal processing algorithms like DOAT are on high demand. We expect this work to impact several fields related to healthcare.

# Methods

**Mathematical description of the DOAT**

Let us assume that we have an optoacoustic signal $P$ with $M$ elements each of them corresponding to a discrete time point $t_n$, $P(t_n) = (p_{t_1}, p_{t_2}, \ldots, p_{t_M})$, where $M$ is uneven. If we multiply the time values by the speed of sound we obtain the signal as a function of the distance to the detector $d'_n$. Moreover, by making a change of variables, using the central point of the signal (($M$-1)/2 + 1) as a reference, we can express the signal as $P(d_n) = (p_{d_{-(M-1)/2}}, \ldots, p_{-d_2}, p_{-d_1}, p_{d_0}, p_{d_1}, p_{d_2}, \ldots, p_{d_{(M-1)/2}})$ where $d_0 = 0$. We can now define a set of DOAT basis functions $\varphi_{d_l}(d_n)$ each of them having the same length as $P(d_n)$. The number of functions contained in the set is (*M-1*)/2 and each function is characterized by the value of $d_l$ as:

$$\varphi_{d_l}(d_n) = \begin{cases} \frac{-1}{\sqrt{2}} & \text{if } d_n = d_l \\ \frac{1}{\sqrt{2}} & \text{if } -d_n = d_l \\ 0 & \text{otherwise} \end{cases} \qquad (1)$$

The values of $d_l$ range from $d_1$ to $d_{(M-1)/2}$. In order to obtain the DOAT coefficients, $C_{d_l}$, one must apply the following formula (the DOAT transform):

$$C_{d_l} = \sum_{n=-(M-1)/2}^{(M-1)/2} P(d_n)\varphi_{d_l}(d_n) \qquad (2)$$

To obtain the original signal from the DOAT coefficients and basis functions we apply:

$$P(d_n) = \sum_{l=1}^{\frac{M-1}{2}} C_{d_l}\varphi_{d_l}(d_n) \qquad (3)$$

**Simulations**

In order to estimate the accuracy of the OA transform vs the DFT and DWT we simulated 10 spheres of diameters between 20µm and 200µm using the formula ([26]):

$$p(r,t) = \frac{r + v_s t}{2r} p_0(r + v_s t) + \frac{r - v_s t}{2r} p_0(-r + v_s t) + \frac{r - v_s t}{2r} p_0(r - v_s t) \quad (4)$$

Where $r$ corresponds to the observation point, $v_s$ corresponds to the velocity of sound in the medium, $t$ represents the time, and $p_0$ is the initial pressure.

We used sampling rates ranging between 3.74GHz and 37.4GHz, decreasing as the radius of the simulated sphere was increased.

We added the effect of a real transducer by convolving each simulated signal with the electric impulse response (EIR) of the simulated transducer. The typical impulse response of a transducer used in RSOM systems consists of a Gaussian filter with a signal capacity range from 10MHz to 120MHz. The filter was applied to each simulated signal using the *gaussianFilter* function from the simulator k-wave[27] , resulting in new signals that showed the effect of a real transducer.

In order to simulate the DC noise, we added a ramp function to each simulated signal. To simulate electrical random noise, we added uniformly distributed random noise bounded by double the maximum and minimum values of the noiseless signal.

Once we applied to DFT, DWT and DOAT to the noisy signals we used specific filters for each transform. For the DFT, we used a 4th order Butterworth bandpass filter, with a low-pass frequency of 10MHz and a high-pass frequency of 120MHz, in order to remove noise outside the detection bandwidth as is standard when processing optoacoustic signals.

For the DWT, we used a Daubechies 8 mother wavelet over 7 decomposition levels. The resulting coefficients were filtered using a soft threshold set by the Rigrsure criteria, following the standard procedure. The Rigsure criteria sets a threshold for each dyadic resolution level of the wavelet coefficients by minimizing the Stein unbiased estimate of risk (SURE) for threshold estimates [28]. With a soft threshold, the wavelet coefficients that lay below the threshold assigned by Rigrsure are proportionally shrunk towards zero, minimizing the noise without completely losing information given by smaller coefficients [29].

Additionally, to remove DC noise, we subtracted the mean of the approximation coefficients, corresponding to the lower frequency components of the signal, from each of the coefficients.

Since there is no general agreement in the literature on what DWT arrangement to apply to optoacoustic signals, we chose all WT processing schemes after assessing the performance of various combinations of WT features used in previous literature. Some of the combinations of mother wavelet, decomposition level and threshold criteria tested were: sym7 with 7 decomposition levels and soft SURE threshold, bior3.5 with 6 decomposition levels and Sqtwolog threshold criteria [30] and Daubechies 8 wavelet with 6 and 7 decomposition levels and Rigrsure threshold criteria [31], [32].

For the DOAT we removed all negative coefficients to account for the constant polarity of optoacoustic waves, and we set to 0 all coefficients corresponding to the basis of spheres with radii larger than 100µm. Finally, we used a moving average filter of length 150 for random noise removal.

In order to compare the processed signals with the original N shaped signal we used the root mean square error (RMSE). RMSE computes the square root of the mean squared error between the original N shaped signal and the processed signal following the formula:

$$RMSE = \sqrt{\frac{1}{n}\sum_{i=1}^{n}|P_i - Q_i|^2} \tag{5}$$

Where $P$ is the original N shaped signal, $Q$ is the processed signal and $n$ is the number of scalar observations. We calculated the RMSE of each of the 10 simulated signals against the three corresponding processed signals. We then obtained the mean RMSE and its standard deviation for each processing scheme (DFT, DWT, DOAT).

**DOAT for optical resolution optoacoustic microscopy data.**

To evaluate the performance of the DOAT against the DFT, we analyzed the Signal-to-Noise Ratio (SNR) of the processed images relative to the raw data and the Root Mean Square Error (RMSE) of the processed images relative to the real image.

The SNR was calculated by selecting specific Regions of Interest within the image, specifically targeting areas of high intensity and background noise. The SNR is defined as:

$$SNR = \frac{<A>}{\sigma} \tag{6}$$

where $A$ is the mean value of the ROI of the image containing signal and $\sigma$ is the standard deviation of a region of the image with only noise.

The metric was calculated across multiple ROIs; the resulting values were then aggregated to determine the mean and standard error.

To further quantify the transform accuracy, the Root Mean Square Error (RMSE) was calculated (Eq. 5). A high-fidelity photographic image served as the ground truth reference (P), against which the reconstructed images (Q) were compared. Following a spatial adjustment to ensure exact pixel to pixel alignment, a histogram matching procedure was applied to the images. This intensity normalization ensured a fair comparison by eliminating global contrast or brightness offsets. Finally, the RMSE was derived from the pixel wise difference between the reference and the reconstructed images.

**DOAT for optoacoustic tomography: the SDOAT.**

The DOAT transform has to be modified to be applied in tomographic tomography obtaining the SDOAT which can be expressed as follows:

$$C_d = \sum P \cdot w_{l,m} \cdot \varphi_d \tag{Eq. 7}$$

Where $w_{l,m}$ is the window function, with length $l$ and shifted $m$ positions around the origin.

All Methods were applied using MATLAB. We used the Wavelet Toolbox to perform WT processing and the k-Wave toolbox to perform the reconstructions.

# Acknowledgements

Dr. Aguirre would like to thank support from the Madrid Autonomous Region Talento Project 2020-T1/TIC-20661.

# Author contributions

J.A conceived the OA transform. J.A develop an initial basic code. M.R extended the code to all the simulations and experiments. M.R and J.A Performed the investigation. A.J Performed the imaging experiments. M.R, A.J and J.A wrote the manuscript. A.J, R.R and S.C. Built the imaging system. J.A Lead and supervised the research. All authors revised the manuscript and have given approval to its final version.

# A physics inspired and efficient transform for optoacoustic systems
# Supplementary Data

María Rodríguez Sáenz de Tejada[1,2], Alvaro Jimenez[1,2], Rodrigo Rojo[1,2], Sergio Contador[1,2], Juan Aguirre[1,2]*

**Supplementary Note 1: Optoacoustic microscopy (OAM) imaging system**

In Suppl. Figure 1, we depict the schematic of the optoacoustic (OA) microscope system. The ultrasound transducer is positioned under the USAF resolution target, with water serving as the acoustic coupling medium between them.

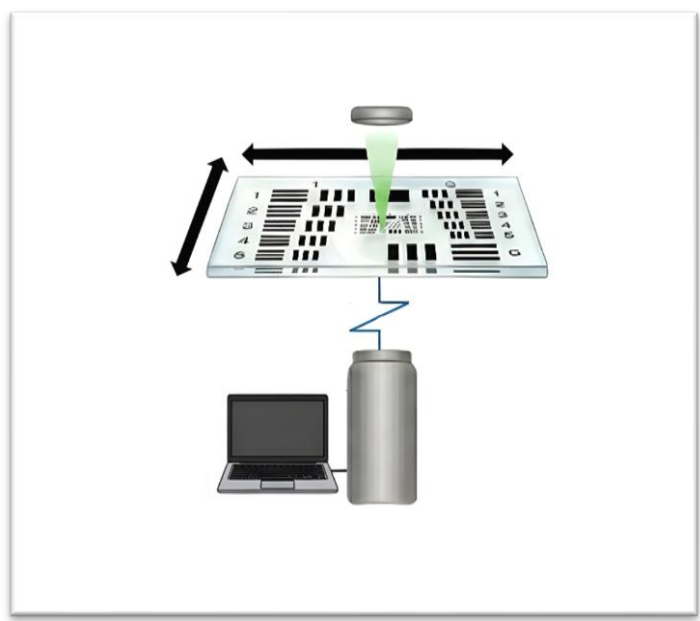

**Supplementary Figure 1:** Schematic layout of the optoacoustic (OA) imaging system.

The illumination source was a Q-switched DPSS laser (Onda NS; Bright Solutions Srl., Cura Carpignano, Italy) with an OEM laser head, controlled by a C-Box micro control unit (Bright Solutions Srl.). The laser is emitted at 532 nm with pulse energy up to 1 mJ, pulse duration of 2–10 ns, and a repetition rate of 4 kHz. The beam was directed through a 25mm diameter Absorptive ND filter (NE05A; Thorlabs, Newton, NJ, USA) and later focused with a 50 mm focal length lens (AC254-050-A; Thorlabs) mounted on a CP33T/M cage plate (Thorlabs) attached to a 50 mm aluminium post (TRA50/M; Thorlabs) and a 40mm post holder (PH40/M; Thorlabs). All of this is vertically aligned by attaching it to a solid aluminium optical breadboard (Thorlabs) placed vertically. This ensures a concentrated energy point for the OR-OAM. For synchronization, a silicon (Si) photodiode (DET10A2; Thorlabs) was used to trigger the DAQ system.

The sample was a positive 1951 USAF Test Target, 3'' x 3'' (R3L3S1P; Thorlabs) placed above a 3D printed water bath, of dimensions 15.24x15.24x2.5 cm (17x17x3 cm outside wall) printed on a LION 2 (LEON3D, Valverde de la Virgen, León, España) with a white filament (White PLA+ 1.75 mm; LEON3D). In the middle in had a 12.5 diameter hole for the transducer. The water bath height can be adjusted with and external stage (XYT1/M; Thorlabs) to which it is mounted.

The movement of the target came from two motorized stages to allow the scanning along two axes: a slow X-axis stage and a fast Y-axis stage. The slow axis was driven by a high-load linear stage with DC servo motor and ball-screw drive (M-414; Physik Instrumente, Karlsruhe, Germany), providing a 100 mm travel range and a maximum velocity of 100 mm/s, controlled by a C-863 Mercury servo controller (Physik Instrumente). The fast axis was driven by a precision linear stage with direct magnetic drive technology (V-408; Physik Instrumente). The fast stage was mounted on the slow stage, with the two stages oriented orthogonally to provide perpendicular motion along the fast and slow scanning axes. This motors where connected to the USAF target through two aluminium Posts with 12.7 mm in diameter. A 75 mm one (TRA75/M; Thorlabs) was mounted to the fast stage subsequently connected to a 50 mm one (TRA50/M; Thorlabs) which in the end was anchored to the USAF target. To have a better weight distribution and to avoid attaching posts to more places which would difficult the scanning, we attached the post to a 3D printed surface, 7.62x2.2x0.3 cm  (White PLA+ 1.75 mm; LEON3D) which we then pasted to the target over its biggest marks, which we were not going to use. It is interesting to note that the point where the two posts are connected is at the same height as the focusing lens.

The optoacoustic signals were detected by an Evident V214-BC-RM high-frequency contact transducer (Evident Scientific, Waltham, MA, USA). This transducer features a 50 MHz central frequency and a 0.25-inch (6.35 mm) element diameter.  The V214 is an unfocused (flat) transducer, providing a collimated beam profile suitable for direct contact measurements. That is the reason why we left the hole, it has the size so that if we surround the transducer with a rubber casing it stays in the water bath and no water flows out. Also, this assures that the signals efficiently travel through the water without any interference. This transducer was coupled with a 5682 preamplifier (Olympus NDT; Waltham, MA, USA) used to amplify the received signal before digitization.

For each capture, the complete scan of one area, optoacoustic signals were digitized with a PicoScope 3406D oscilloscope (Pico Technology, St Neots, UK) at a sampling rate of 250 MS/s within a 5 µs acquisition window. The recorded waveforms were streamed to a laptop for processing and visualization using custom MATLAB and PYTHON routines.